\documentclass{aastex}
\usepackage{spr-astr-addons}
\usepackage{url}

\RequirePackage{color}

\newcommand{\emaila}{bltan@nao.cas.cn}

\begin{document}

\title{A Physical Explanation on Solar Microwave Zebra Pattern with the Current-carrying Plasma Loop Model}
%\slugcomment{Not to appear in Nonlearned J., 45.}
%% Running heads
\shorttitle{Explanation on Solar Radio Zebra Pattern}
\shortauthors{Baolin Tan}

\author{Baolin Tan\altaffilmark{1}}

%\affil{Key Laboratory of Solar Activity, National Astronomical
%Observatories, Chinese Academy of Sciences, Datun Road A20,
%Chaoyang District, Beijing 100012, China}

\email{\emaila}

\altaffiltext{1}{Key Laboratory of Solar Activity, National
Astronomical Observatories, Chinese Academy of Sciences, Beijing
100012, China. email: bltan@nao.cas.cn}

%\author{Feng Liu\altaffilmark{2}}
%%\affil{Southwestern Institute of Physics, Chengdu 610041, China.}
%\altaffiltext{2}{Southwestern Institute of Physics, Chengdu
%610041, China.}

\begin{abstract}

Microwave zebra pattern structure is an intriguing fine structure
on the dynamic spectra of solar type IV radio burst. Up to now,
there isn't a perfect physical model for the origin of the solar
microwave zebra pattern. Recently, Ledenev, Yan and Fu (2006) put
forward an interference mechanism to explain the features of
microwave zebra patterns in solar continuum events. This model
needs a structure with a multitude of discrete narrow-band sources
of small size. Based on the model of current-carrying plasma loop
and the theory of tearing mode instability, we proposed that the
above structure does exist and may provide the main conditions for
the interference mechanism. With this model, we may explain the
frequency upper limit, the formation of the parallel and
equidistant stripes, the superfine structure and intermediate
frequency drift rate of the zebra stripes. If this explanation is
valid, the zebra pattern structures can reveal some information of
the motion and the inner structures of the coronal plasma loops.

\end{abstract}
\keywords{Solar microwave emission, Zebra pattern structure,
electric current loop, flare}

%\section*{}
%\label{sec:intro}

\section{Introduction}

During much of solar flares, it is very frequently found that a
kind of intriguing fine structure pattern superposed on the solar
radio broadband spectrum of type IV bursts, and behaved as a
series of almost parallel and equidistant stripes in the dynamic
spectrum. Such structure is called zebra pattern. Most often,
zebra patterns are observed in meter and decimeter frequency
range, and with up to 10 and more stripes (Slottje, 1972). It is
seldom to observe zebra patterns in microwave frequency range in
the early observations, and even if we found them, there always
only 3 or 4 stripes in a zebra pattern structure (Ning et al,
2000; Ledenev, Yan and Fu, 2001). However, in recent microwave
observations, some remarkable zebra patterns are also found with
up to 30 stripes in the frequency range of 2.60 -- 3.80 GHz
(Chernov et al, 2005). Fig. 1 is an example of zebra pattern
occurred in the frequency of 2.90 -- 3.80 GHz observed at Chinese
Solar Broadband Radiospectrometer (SBRS/Huairou) in 02:42:55 --
02:43:20 UT, 13 Dec. 2006 in the famous flare event, with strong
right polarization, 5 stripes, and the duration is about 10
seconds, cognizably. Up to now, the upper limit frequency of Zebra
pattern structure is below 6 GHz, and the corresponding wavelength
is about 5 cm (Altyntsev et al, 2005). The duration of zebra
pattern event is from several to a few decades of seconds. It is
uncommon to last for more than 30 seconds.

\begin{figure}
\begin{center}
 \includegraphics[width=8cm]{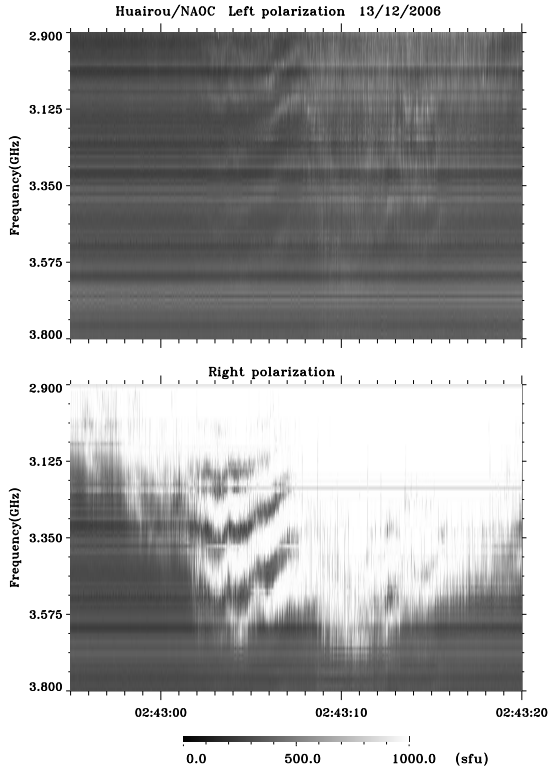}
  \caption{An example of Zebra pattern structure occurred in the frequency of 2.90 -- 3.80 GHz observed at Chinese Solar
  Broadband Radiospectrometer (SBRS/Huairou) in 02:42:55 -- 02:43:20 UT, 13 Dec. 2006.
  The upper and lower panels are left and right polarization components, respectively.}
\end{center}
\end{figure}

If we fix the time and plot the profile of the emission flux with
respect to frequency in the Zebra pattern structure, we may find
that the flux profile behaves periodic feature, and the period is
the frequency gap between two stripes. Fig. 2 gives an example
profile of the emission flux at 02:43:05 UT, 13 Dec. 2006 of the
Zebra pattern showing in Fig.1. There are 5 peaks in this profile,
and each represents one stripe.

\begin{figure}
\begin{center}
 \includegraphics[width=8cm]{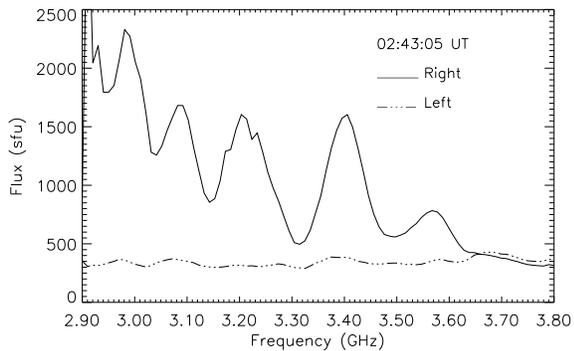}
  \caption{A profile of the zebra pattern emission flux with respect to the frequency at a fixed time of 02:43:05 UT, 13 Dec. 2006.
  The solid and the dash-ploted curves are indicated the right and left polarization components, respectively.}
\end{center}
\end{figure}

The another main feature of Zebra pattern structure is the
intermediate frequency drift against the continuum emission of
type IV radio outbursts (Slottje, 1972; Mollwo, 1983; etc). Fig. 1
shows that frequency drift rate is about 55 MHz/s during 02:42:59
-- 02:43:03 UT, and -45 MHz/s during 02:43:03 -- 02:43:10 UT
around the frequency of 3.50 GHz, negatively or positively.

Chernov et al (2003, 2008) have found that the Zebra pattern
stripes in the microwave range often have some superfine
structures, consisting of separate spike-like pulses with
millisecond duration. Chen and Yan (2007) also found that Zebra
stripes consist of periodically narrow band pulsating superfine
structures, and the period is about 30 milliseconds. Because of
the saturation around the center of the Zebra stripes associated
with the flare event occurred in 13 Dec. 2006 (see in Fig. 1), we
could not distinguish the obvious superfine structures in this
event. However, from the limb parts of the Zebra stripes, we may
also find some evidences of the quasi-periodic narrow band
pulsating superfine structures. Fig. 3 shows the quasi-periodic
narrow band pulsating superfine structures of the Zebra stripe in
a segment of 02:43:02.8 -- 02:43:03.8 UT at frequency of 3.65 GHz,
near the limb of the fourth stripe (numbered from low frequency to
high frequency). And the period of the pulsating superfine
structures is about 30 -- 35 milliseconds.

\begin{figure}
\begin{center}
 \includegraphics[width=8.cm]{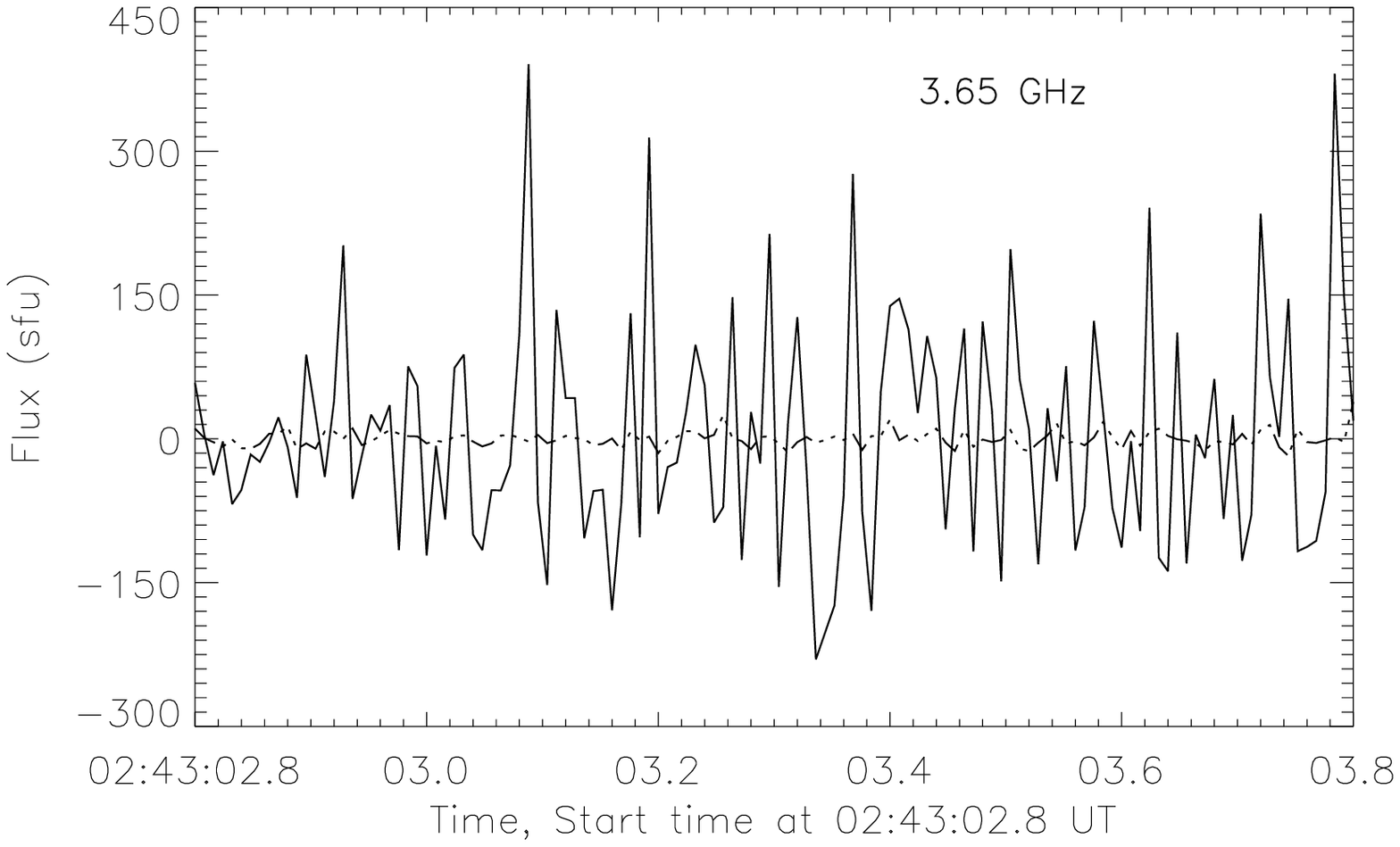}
  \caption{The quasi-periodic narrow band pulsating superfine structures of the zebra stripe in a segment of 02:43:02.8 -- 02:43:03.8 UT at
  frequency of 3.65 GHz, near the limb of one stripe.}
\end{center}
\end{figure}

Generally, people think that the microwave Zebra pattern may
provide some useful information about the kernel of the flaring
regions. It is necessary for any model of zebra pattern to explain
the following features: (1) the upper limit of frequency of Zebra
pattern, (2) almost parallel and equidistant stripes, (3)
superfine structures, (4) intermediate frequency drift rate. In
order to interpret the formation of Zebra pattern, a great number
of theoretical models were proposed. These models can be
classified simply into two groups:

(1) Isogenous models, which proposed that all the stripes in a
Zebra pattern are generated from a single emission source, and the
emission mechanism is a kind of nonlinear coupling between two
Bernstein modes, or Bernstein mode and some electrostatic upper
hybrid mode waves. The frequency gap between two stripes is very
close to the frequency of electron cyclotron emission (Rosenberg,
1972; Chiuderi et al, 1973; Zaitsev and Stepanov, 1983).

(2) Heterogenous models, which proposed that all the stripes in a
Zebra pattern are generated from different emission source
regions, and different source region are located at different
positions in the magnetic flux tube. In each source region, some
resonant conditions are satisfied, and the emission mechanism is
probably the coupling between whistler wave and electrostatic
upper hybrid mode wave which triggered by some plasma
micro-instabilities. The frequency gap between two stripes is
greatly different to the frequency of electron cyclotron emission
(Kuijpers, 1975; Fomichev and Fainshtein, 1981; Mollwo, 1983;
Ledenev, Yan, and Fu, 2001; Chernov et al, 2005; Altyntsev et al,
2005).

However, up to now, there is no perfect theoretical model which
can explain Zebra pattern satisfactorily. Recently, Ledenev, Yan,
and Fu (2006) proposed that Zebra pattern is possibly formed from
some interference mechanism in the propagating processes. They
assumed that there are some inhomogeneous layers with small size
in solar coronal plasma, and such structure will change the radio
waves into direct and reflected rays. When the direct and
reflected rays meet at the position of observer, interference will
take place and form Zebra pattern structure.

When the dimensional size of an emission source region is smaller
than the minimum wavelength of emission spectrum, the source
region can be treated as a point source. At the same time, when
the emission spectrum is continuum, then it is possible to form an
interference pattern. The dimensional size of the source region is
much smaller than the characteristic size of plasma density
gradient, and can be regarded as a point source. However, the
dimensional size of the source region is always much larger than
the emission wavelength. The interference condition requires that
the source region has a narrow, zonal inhomogeneous interior
structure to keep some definite phasic difference between the
direct and the reflected rays. At the same time, in order to
generate definite interference strength, the number of the point
sources should be abounded. Then, is there plenty of such abounded
point sources in the solar flaring region? What mechanism can
generate such structures?

In fact, the abounded point sources are necessary not only in the
interference model of Ledenev, Yan, and Fu (2006), but also in
other zebra pattern models. For example, the model of LaBelle et
al (2003), in which assumed that the emission is generated from
the double plasma resonant layer in coronal loop. The presence of
localized density irregularities within the type IV source region
leads to trapping of the upper hybrid Z-mode waves in density
enhancements, transforms into electromagnetic waves by the
electron-cyclotron maser mechanism, and forms the zebra pattern.
In such case, the localized density irregularity is a necessary
condition, and a number of point sources can meet this condition
naturally.

Briefly, the structure with abounded point sources is an important
condition for the formation of zebra pattern. Based on the
analysis of current-carrying plasma loop model and the related
resistive tearing-mode instability, this work proposed that the
tearing mode magnetic islands in the current-carrying plasma loop
can form a reasonable structure to generate the interference
mechanism and produce the Zebra patterns. We introduce the
formation of the tearing-mode magnetic islands in current-carrying
plasma loop in section 2. Then in section 3, we present a detailed
explanation of the interferential rays and the formationtions of
Zebra patterns. At the final, some summaries and discussions are
given in section 4.

\section{Current-carrying Plasma Model and Magnetic Island}

There are much of evidences showing that the solar flaring region
is always composed with many magnetic flux tubes, and the magnetic
flux tubes are always current-carrying plasma loops (Alfven \&
Carlqvist, 1967; Melrose, 1991, 1995; Ashbourn \& Woods, 2004;
Tan, 2007; etc). In such loops, the magnetic field can be
decomposed into three components: (1) longitudinal component
$B_{\varphi}$ generated from the convection motion of photosphere
or sub-photosphere, (2) poloidal component $B_{\theta}$ induced by
the electric current flowing along the plasma loop, and (3) the
radial component $B_{r}$ which is a disturbed quantity. At the
equilibrium state, $B_{r}\simeq 0$. Usually, we may define a
safety factor to describe the equilibrium property of the
current-carrying plasma loops:
$q(r)=\frac{arB_{\varphi}}{RB_{\theta}}$, $r$ is the distance to
the axis of the loop (generalized to the section radius $a$), $R$
is the loop radius. When $q(r_{s})=\frac{m}{n}$, and $m$ and $n$
are positive integers, then it defines a rational surface, $m$ and
$n$ represent the poloidal and toroidal mode number, respectively,
$r_{s}$ defines the position of the rational surface. Different
$m$ and $n$ define different rational surfaces. A series of
rational surfaces form a coaxial nested configuration (see the
Fig.1 and Fig.2 of Tan and Huang, 2006). Between two rational
surfaces there are countless irrational surfaces where the safety
factor couldn't be expressed as a ratio of two positive integers.

In the above current-carrying plasma loops, the magnetic field
lines are helical along the longitudinal direction. There will do
exist magnetic shearing between the neighboring rational surfaces
with different radius. When the plasma has finite resistivity, the
magnetic shear will easily trigger the resistive tearing-mode
instability and its evolution (Furth, Rutherford and Selberg,
1973; Specer, 1977; Tan \& Huang, 2006). While such process
occurs, the magnetic reconnection will take place between the
neighboring rational surfaces, and the regular rational surfaces
will evolve into a series of magnetic islands. These magnetic
islands are distributed spirally along the longitudinal rational
surface (left panel in Fig. 4), and behaves like a series of
convex mirrors in the three-dimensional space (right panel in Fig.
4). In the inner of the magnetic island, the plasma density
increases from the limb to its core, continuously. From the global
view, a multiple of magnetic islands distributed like a crystal
lattice in the plasma loop.

\begin{figure}
\begin{center}
 \includegraphics[width=7.0cm]{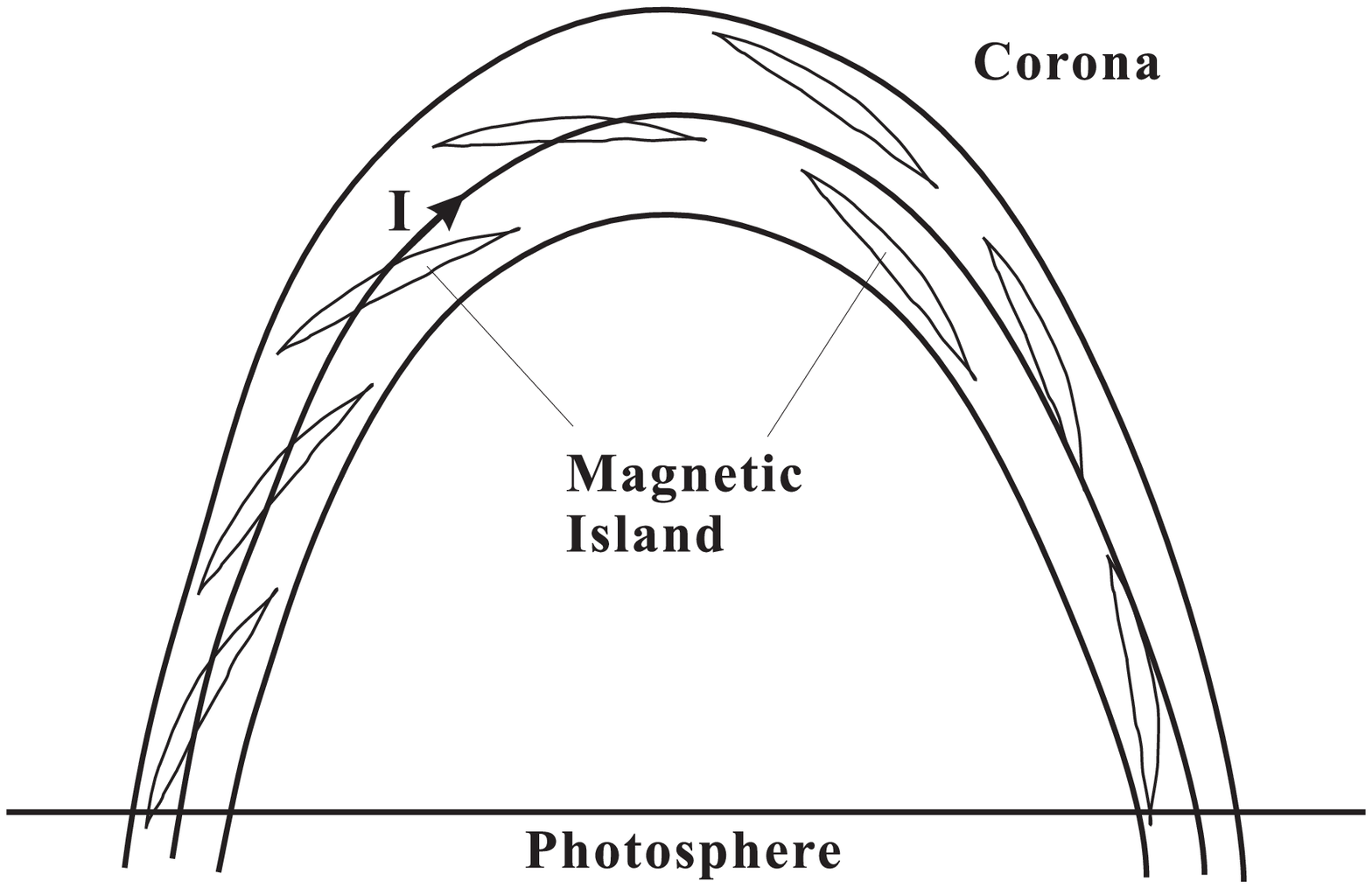}
 \includegraphics[width=6.5cm]{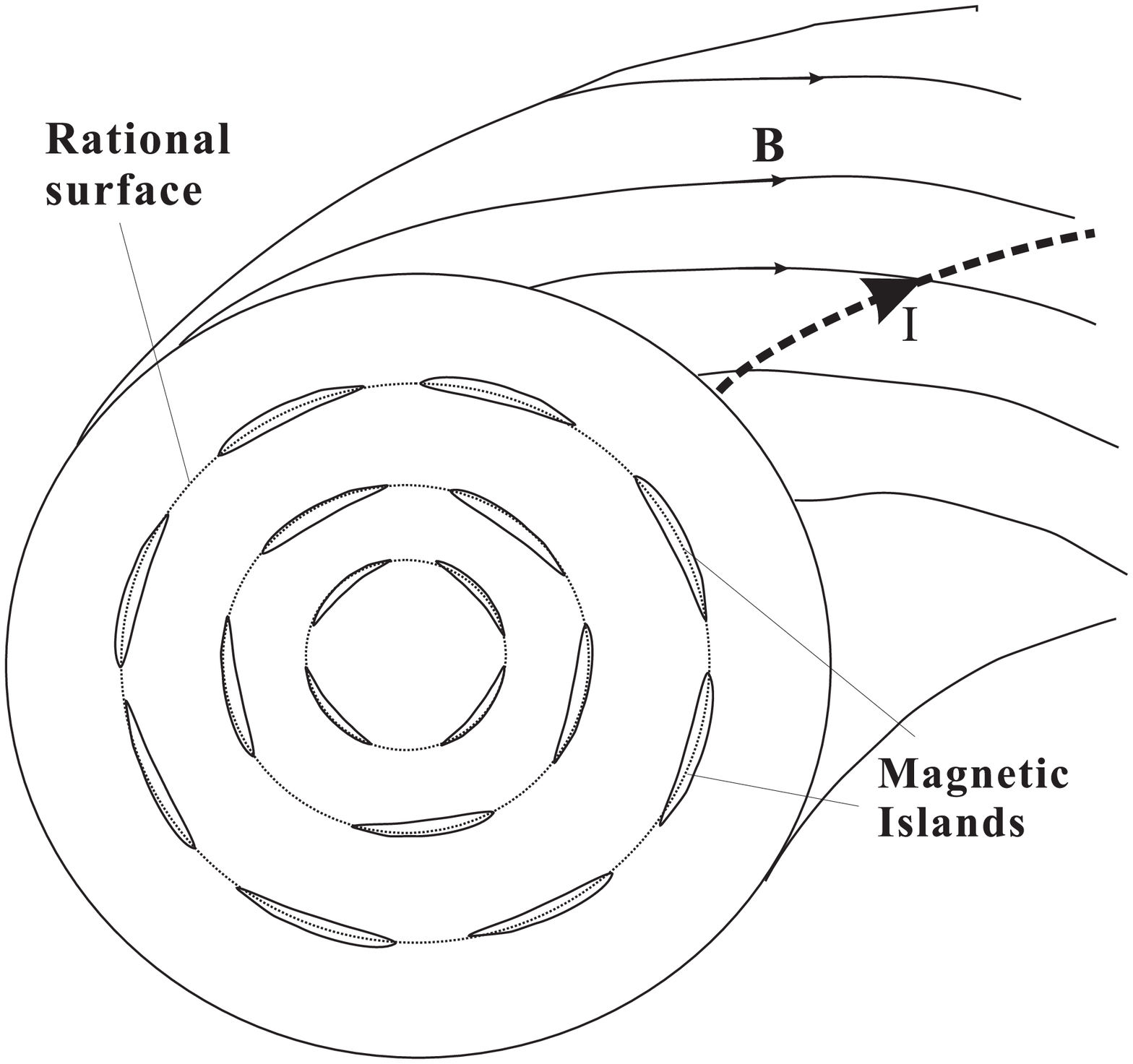}
  \caption{The schematic view of the spatial distribution of the magnetic islands in the current-carrying plasma loop generated from the tearing mode
  instabilities. Here, the size of the magnetic islands is not plotted by the real scales. The upper panel is a longitudinal projection,
  and the lower panel is a transverse projection in the loop's section in which a few magnetic field lines ($\mathbf{B}$) and electric current ($\mathbf{I}$) are drawn. }
\end{center}
\end{figure}

According to the nonlinear tearing-mode equations, we may obtain
the width of the magnetic island:
$w(r_{s})=4a(\frac{r_{s}qB_{r1}}{mq'B_{\theta}})^{1/2}$, here
$q'=(dq/dr)_{r=r_{s}}$, $B_{\theta}$ is mainly dominated by the
longitudinal electric current $I$, $B_{r1}$ is the disturbed
radial magnetic field. Assume that the distribution of the current
density in the cross section is in the form of pinch regime
(Bennett, 1934): $j=j_{0}e^{-r^{2}}$, the total current is $I$,
$j_{0}=eI/[(e-1)\pi]$. Then we may obtain the poloidal magnetic
field as:

\begin{equation}
B_{\theta}(r)=\frac{\mu_{0}}{r}\int_{0}^{r}
j(x)xdx=\frac{\mu_{0}eI}{2\pi(e-1)r}(1-e^{-r^{2}}).
\end{equation}

Substitute the above relation into the expression of the magnetic
island width:

\begin{equation}
w(r)\approx5.03\times10^{3}(\frac{r^{3}B_{r1}(1-e^{-r^{2}})^{2}}{mI(1-e^{-r^{2}}-r^{2}e^{-r^{2}})})^{1/2}a.
\end{equation}

Here, $r$ is generalized radial parameter with respect to the
section radius $a$. From Equation (2), we know that the magnetic
island width is mainly dominated by the rational surface radius
($r$), total current ($I$), and the disturbed radial magnetic
field ($B_{r1}$). If we suppose: $B_{r1}=1$ Gs, $I=10^{11}$ A,
$a=10^{7}$ m, when $r=1$ (at the surface of the loop), $w\simeq
1.4\times 10^{3}$ m; when $r=0.5$, $w\simeq 3.8\times 10^{3}$ m.

According to the theory of resistive tearing-mode instability, the
magnetic reconnection is mainly generated from the vicinity of the
separatrix and will produce an induced electric field paralleled
to the magnetic field (Apicer, 1977; Kuijpers et al, 1981; etc.).
With this induced electric field the particles can be accelerated
near this place and drive the plasma emission to radiate (Drake et
al, 2006; Karlicky and Barta, 2007). Then, what is the size of the
emission source? From the dynamic analysis of the nonlinear
tearing-mode, we cam obtain the width of the separatrix which is
in the same order as the thickness of the magnetic island:

\begin{equation}
\delta\approx5.6a\times10^{-3}(\frac{\gamma\eta\rho}{m^{4}B_{\varphi}^{2}}\cdot\frac{r^{4}}{(2r^{2}e^{-r^{2}}+e^{-r^{2}}-1)^{2}})^{\frac{1}{4}}.
\end{equation}

Here, $\gamma$ is the growth rate of the tearing-mode instability,
$\eta$ plasma resistivity, and $\rho$ plasma density. We need to
note that the above result is deduced from the assumption of
$j=j_{0}e^{-r^{2}}$, and the corresponding plasma density is also
concentrate to the axis of the current-carrying plasma loop with
decreasing along the section-radius. We may approximately assume
that $\rho=\rho_{0}e^{-r^{2}}$. Then we may let $B_{\varphi}=500$
Gs, $n_{i}=10^{16}$ m$^{-3}$, $T_{e}=500$ eV, and find that the
thickness of the magnetic island is about $\delta\simeq 8$ cm at
the loop surface , and $\delta\simeq 5$ cm near the center of the
loop. Furthermore, when we change the values of the plasma
parameter and magnetic field strength, we find that the $\delta$
value is always kept in order of decimeter, and the corresponding
frequency is $\leq 6.0$ GHz.

\section{Formation of interferential rays and explanation of Zebra pattern}

\subsection{Interference Process}

The great number of tearing-mode magnetic islands form a crystal
lattice-like structure in the plasma loops. Around each X-point,
there will be an induced electric field, the electrons will be
accelerated around such place (Drake et al, 2006; Karlicky and
Barta, 2007). And the energetic electrons will produce some plasma
turbulence, and generate plasma emission (Dulk, 1985) near the
magnetic islands. From the above analysis, we may obtain the
conclusion that the width of the magnetic islands is about several
kilometers which is longer than the wavelengths of metric,
decimetric, and centimeter waves, and at the same time the
thickness of the magnetic island is about several centimeters
which is shorter than the wavelengths from metric to centimeter
waves. So the microwave emission source can be considered as a
point source. It is narrow-band emission. When the emission
propagates from the place near the inner island to the outward, it
may meet the outer islands and decompose into two rays, one is the
direct ray, and the other is reflected ray through the island.
There will have a phase difference between the two rays when they
come to the observer. Then they will interfere with each other and
form an interference structure. According to the work of Ledenev
et al (2006), such interference structure will behave as the Zebra
pattern structure. Because of the assumption of pinch regime, the
plasma density decreases from the inner to the outer of the loop.
As a result, the frequency of the plasma emission also decreases
from the inner part to the outer of the loop. So, the emission can
escape from the loop, and propagate to the observer.

\begin{figure}
\begin{center}
 \includegraphics[width=7.5cm]{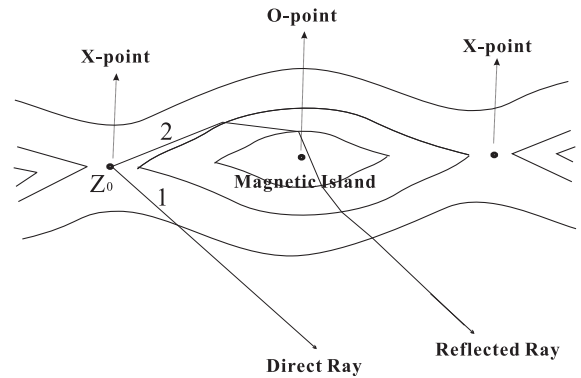}
  \caption{Trajectories of the direct and reflected rays formed from the magnetic islands in the current-carrying plasma loop.}
\end{center}
\end{figure}

Fig. 5 presents the trajectories of the direct and reflected rays
formed from the tearing-mode magnetic islands, both of beam 1 and
beam 2 come from $Z_{0}$. Beam 1 is a direct ray which does not
pass through the magnetic island, while beam 2 is a reflected rays
which enters the region of magnetic island, undergoes a series of
refractions and comes out from the other side of the magnetic
island. There will be a phase difference between beam 1 and beam 2
after they run through the island. They will interfere with each
other and form a Zebra-like patterns when they meet and be
observed by the ground-based radio telescopes.

\subsection{Explanation of the Zebra pattern}

(1) The upper limitation of frequency of Zebra pattern

From equation (3), we find that when $B_{\varphi}=500$ Gs,
$n_{i}=10^{16}$ m$^{-3}$, $T_{e}=500$ eV, the thickness of the
magnetic island is in the range of $\delta\simeq 5 - 8$ cm. Even
if we change the values of the parameters, we always get $\delta
\geq 5$ cm. If the above interference model is valid, then it is
reasonable that zebra patterns only occur in the radio emission
with frequency lower than 6 GHz, and doesn't emerge from the
observations of millimeter wave or infrared emission, because the
wavelength of the latter is much shorter than the thickness of the
magnetic islands, the source region couldn't be regarded as a
point source.

(2) the formation of the Zebra stripes

Generally, the microwave emission frequency is mainly related to
the magnetic field strength, temperature and plasma density in the
source region. Because the size of the magnetic island is very
small, the variation of the magnetic filed strength and
temperature around the magnetic island in the current-carrying
plasma loop is not obvious, we may neglect its effect on the
microwave emission frequency. Then the main factor affected to the
frequency is the plasma density in the source region. If the
emission mechanism of the Zebra pattern structure is mainly the
plasma mechanism, the relation between the frequency $f$ (unit in
Hz) and the plasma density $n_{e}$ (unit in m$^{-3}$) is
$f=sf_{pe}\simeq 9sn_{e}^{1/2}$, $f_{pe}$ is the plasma frequency,
and $s$ the harmonic number. Consisting with the assumption of the
distribution of the current density in the cross section, we may
also assume that the distribution of the plasma density is in the
form of $n_{e}=n_{0}e^{-r^{2}}$. Then the plasma density decreases
from the center of the loop to loop surface. The emission
frequency will also decrease from the center of the loop to the
loop surface. However, from the above analysis, we know that the
Zebra pattern emission is mainly generated from the region around
the magnetic islands, which is not distributed continuously, but
localized along the rational surface. As a consequence, the
emission frequency is not continuous. When the emission is come
from the island region with higher density, the frequency is
higher. When it comes from the region between two rational
surfaces with lower density, then the frequency is lower, and the
spectrum presents stripe structures. One stripe represents the
emission produced from one rational surface, and different stripes
come from different rational surfaces. Fig.2 shows that the Zebra
pattern emission flux behaves periodic feature, the period is the
frequency gap between two stripes which reflects the difference of
the plasma densities between two rational surfaces. If this
deduction is valid, we may investigate the inner structure feature
of the coronal plasma loop by studying Zebra pattern structures.

(3) the superfine structure of the Zebra stripes

In the work of Chen and Yan (2007), they explained the superfine
quasi-periodic structure by using relaxation oscillations, which
modulate the electron cyclotron maser emission that forms the
Zebra stripes during the processes of wave-particle interaction
generated by loss-cone instability of trapped electrons under
double plasma resonance (DPR) conditions (Winglee and Dulk, 1986).

However, in the Zebra pattern interference model, the emission is
supposed as plasma mechanism with narrow band, the relaxation
oscillation is not the suitable mechanism for the superfine
structures. From the work of Tan et al (2007) and Tan (2008), we
know that the current-carrying plasma loop can drive the
tearing-mode oscillations and modulate the microwave emission to
form quasi-periodic pulsations with low down to about 30
millisecond periods. Then, it is reasonable to explain that the
tearing-mode oscillation of the current-carrying plasma loop can
also modulate the emission of the Zebra pattern stripes and form
the superfine structures with some millisecond quasi-periodic
pulsating features.

(4) the frequency drift

From the above analysis we know that the frequency of Zebra
pattern emission is closely related to the plasma density of the
current-carrying plasma loop, then all the variations of the
plasma density will change the emission frequency and result in a
frequency drifting rate. As the order of frequency drifting rate
of Zebra pattern is similar to that of the global frequency
drifting rate of microwave pulsating structures (Tan et al, 2007),
and in that work the global frequency drifting rate of microwave
pulsating structures was explained as the motion of the
current-carrying plasma loop with respect to the ambient coronal
plasmas. Similarly, we may also suppose that the frequency drift
rate of the zebra stripes reflects the motion of the
current-carrying plasma loop. When the loop moves upwards, then
the plasma density decreases generally with time, and the emission
frequency will drift from higher to lower, the drifting rate is
negative; if the loop moves downwards (for example, shrinkage),
the plasma density increases with time, and the emission frequency
will drift from lower to higher, the drifting rate is positive.

Based on the assumption of plasma emission mechanism, we have
$f=sf_{pe}\simeq 9sn_{e}^{1/2}$, then the frequency drifting rate
can be estimated as:

\begin{equation}
\frac{df}{dt}\simeq\frac{9s}{2n_{e}^{1/2}}\frac{dn_{e}}{dr}\frac{dr}{dt}=\frac{f}{2H}v.
\end{equation}

Here, $H=n_{e}/\frac{dn_{e}}{dr}$ is the barometric scale height.
$v=\frac{dr}{dt}$ is the moving velocity of the plasma loop. In
most cases $H\sim 10^{4}$ km. From the introduction of the example
of Fig.1, we know that frequency drift rate of the Zebra stripes
is about 55 MHz/s during 02:42:59 -- 02:43:03 UT, and -45 MHz/s
during 02:43:03 -- 02:43:10 UT around the frequency of 3.50 GHz.
Substitute these values into equation 4, we may find that the
moving velocity of the plasma loop is about 315 km/s downwards
during 02:42:59 -- 02:43:03 UT and 257 km/s upwards during
02:43:03 -- 02:43:10 UT. Here, we simply neglect the effects of
the geometrical projections.

At the same time, it is necessary to note that the moving velocity
of the plasma loop is proportional to the barometric scale height:
$v=2H\cdot \frac{1}{f}\frac{df}{dt}$. In the flaring region, the
magnetic configuration becomes very complex, and the barometric
scale height $H$ will become smaller than the general cases. Then
from Equation (4), we find that the real moving velocity of the
plasma loop may become smaller than above estimated values.

With the above assumption of magnetic field and the plasma
density, we may get the Alfven velocity is about $3.5\times10^{3}
\sim 1.1\times10^{4}$ km/s, which is much faster than the velocity
of the current-carrying plasma loops estimated above. So we
believe that the motion of the current-carrying plasma loops is
only a kind of global motion of which is very slow, and its driver
may not be the magnetic interaction. Possibly, it should be
related with the convection motion below the atmospheric plasmas.

(5) duration of the Zebra pattern

The previous observations show that the durations of Zebra pattern
is in the range of from several seconds to one or two decades
seconds. It is most uncommon to distinguish a Zebra pattern with
over 30 seconds of duration. From our above analysis, the Zebra
pattern structures is possibly formed by a interference mechanism
from a great number of tearing-mode magnetic islands. Then the
duration of the Zebra pattern structures will approximate to the
duration of the resistive tearing-mode instability: $D\leq
0.1513(\frac{\Delta'}{jm\dot{B}_{\theta}})^{2/3}t_{A}^{2/3}t_{r}^{1/3}$
(Tan and Huang, 2006; Tan, 2008). By substituting the above
parameters, we may get: $D\sim 200 - 300$ s. However, it is only
when the plasma loop is at the best state, the duration can last
for such a long-term. Actually, as the flaring region is very
complex, and there will exist many kinds of interferences between
the different loops, the Zebra pattern coming from one plasma loop
is most frangible by the emission from other loops. So the real
duration will be always shorter than the estimated values.

\section{Summary and Discussion }\label{sec:concl}

From the above analysis and estimations, we obtained the following
conclusions: the interference model can give a reasonable
explanation of the radio emission with Zebra pattern structures,
and the current-carrying plasma loop model can provide all the
necessary conditions for the interference model, and can be
applied to explain the other features, such as the the superfine
structure, the upper frequency limit, the intermediate frequency
drifting rate of the zebra stripes, and the durations. If this
model is really valid, the zebra pattern can provide an useful
tool for studying the current-carrying plasma systems in the solar
flaring region. From the scrutinizing of the microwave Zebra
pattern structures, we may get much of information about the
motion and the inner structure feature of the plasma loop, and can
reconstruct the space configuration of the emission source region,
because it is believed that they are closely related to the flare
primary-energy releasing processes. However, there are much of
works need to do theoretically and observationally.

On the other hand, the large numbers of magnetic islands inside
the plasma loops can provide much more opportunities to accelerate
the large numbers of energetic particles in flare events.
According to the previous works (Miller et al, 1997; Vlahos,
Isliker \& Leperti, 2004; etc), the number of accelerated
electrons with energy above 20 keV is roughly $10^{34} - 10^{37}$
electrons $s^{-1}$ in a flare event, and in some X-class flares
the number can be reached to $10^{37}$ electrons $s^{-1}$. The
current-carrying plasma loop model indicates that the particle
accelerations may occur not only in the cusp configuration above
the coronal loop, but also can take place inside the whole loop.
With this model, because the total volume of magnetic islands is
much lager than that of the current sheet near the cusp
configuration above the loop, and the plasma is denser in the
loops, there are much more electrons which can be accelerated by
the induced electric field generated from the tearing magnetic
reconnections. As for the number of magnetic islands associated
with the accelerated electrons, we may give a roughly estimation:
the width of the magnetic island is about 2 km, thickness about 5
cm, length can be assumed as 1\% of the loop's length (assumed as
about $10^{5}$ km). Then the volume of one island is in the order
of $10^{14}$ cm$^{3}$. The plasma density could be assumed
$10^{10}-10^{11}$ cm$^{-3}$. Then the number of accelerated
electrons around one island is about $10^{24}-10^{25}$. In order
to meet the number of $10^{37}$ electrons s$^{-1}$ accelerated in
big flares, the number of islands should be $10^{12}-10^{13}$. If
we suppose the number of flux tubes in the flaring region is about
10 -- 100, then there are about $10^{10}-10^{11}$ magnetic islands
in one plasma loop. We may find that the volume of all islands is
only a small fraction ($\sim$ 0.1\%) in the flaring region. In
fact, it was a big problem to explain the number of $10^{37}$
electrons s$^{-1}$ accelerated in some X-class flares perfectly.
However, in this work, it is not our main task to investigate the
particle acceleration, the above discussion is only a roughly
estimation.

Excited by these energetic particles accelerated around the
tearing-mode magnetic islands in the current-carrying plasma
loops, a series of fine structures will be formed in the microwave
spectrograms. In fact, by using this model, we explained the fast
quasi-periodic pulsations occurred in the famous flare event of 13
Dec. 2006 (Tan et al, 2007). This work is an another attempt.
However, it needs to study in more detailed.

\acknowledgments

The author would like to thank the anonymous referee very much for
the helpful and valuable comments on this paper. This work was
supported by NSFC Grant No. 10733020, 10873021, CAS-NSFC Key
Project (Grant No. 10778605), and the National Basic Research
Program of the MOST (Grant No. 2006CB806301).

\label{}

\end{document}